# Understanding emotions in the context of IT-based self-monitoring

Danielly de Paula*, Florian Borchert, Ariane Sasso, Falk Uebernickel

Hasso Plattner Institute, Potsdam, Germany

**Abstract**

This study explores the intersection of information technology-based self-monitoring (ITSM) and emotional responses in chronic care. It critiques the lack of theoretical depth in current ITSM research and proposes a dynamic emotion process theory to understand ITSM's impact on users' emotions. Utilizing computational grounded theory and machine learning analysis of hypertension app reviews, the research seeks to extend emotion theory by examining ITSM stimuli and their influence on emotional episodes, moving beyond discrete emotion models towards a continuous, nuanced understanding of emotional responses.

## 1. Introduction

The growing interest in information technology-based self-monitoring (ITSM) for chronic care such as mobile apps, and affordable sensing wearables improve the capabilities of SM-based healthcare programs (Jiang and Cameron 2020). ITSM is an emerging medical approach that offers great potential in the self-management of chronic diseases through the involvement of individuals in the self-monitoring of behavioral and biological information (Hui et al. 2019). For instance, features concerning information exchange, health literacy and peer support without the constraints of time and geography helps patients better understand their health patterns over time (K. Liu, Xie, and Or 2020).

\_\_\_\_\_\_\_\_\_\_\_\_\_\_\_\_\_\_\_\_

*Corresponding author:  danielly.depaula@hpi.de

The importance of ITSM is evidenced by the growing number of healthcare consumers who purchase wearable devices that are connected with health apps and the explosion of health apps in the market (Jiang and Cameron 2020).

However, studies on ITSM have been recently criticized for the paucity of strong theory (Jiang and Cameron 2020; Chatterjee et al. 2021). Current studies have mostly focused on investigating the effectiveness of health interventions (K. Liu, Xie, and Or 2020; Guo et al. 2020) rather than contributing to theory, with a few exceptions (Spruijt-Metz et al. 2022; Thompson and Whitaker 2020; Savoli, Barki, and Pare 2020; Brohman et al. 2020).

Although understanding the link between ITSM use and patient outcome is fundamental for healthcare research, it is important to realize the potential that information systems (IS) research can offer through the theoretical understanding of ITSM use and its effects. In particular, IS scholars have highlighted the need to investigate the underlying causal mechanisms of ITSM and users' emotional expressions (Jiang and Cameron 2020; Chatterjee et al. 2021) - as at the same time, emotion theorists have started to question how recent technological advances impact emotional responding (Nozaki and Mikolajczak 2022; Yu et al. 2023).

Emotions are not random events – they occur when something relevant happens to us. Emotion is conceptualized as "an emergent, dynamic process based on an individual's subjective appraisal of significant events." (Scherer 2009) In the context of the theory of emotion process (Nico H. Frijda 1986; N. H. Frijda 1988), *emotions are generated* as reactions to stimuli that are *appraised* to be relevant to the needs and goals of an individual, and once activated, emotions form *action tendencies* (Nico H. Frijda 1986; Moors 2009). Emotions are also

*regulated* to facilitate how individual's experience and express these emotions, which is essential for optimal functioning and well-being (Sheppes, Suri, and Gross 2015; Augustine and Hemenover 2009). While the importance of understanding emotions in IS research has been recognized (Beaudry and Pinsonneault 2010; Liang et al. 2019; Gregor et al. 2014), progress on the topic is constrained due to two main shortcomings.

First, previous work has mainly investigated the relation between specific appraisal patterns and specific emotions, using self-report methods as their primary source (Vornewald, Eckhardt, and Krönung 2015). However, the use of self-report for investigating emotion causation has been the target of severe criticism by theorists in psychology (Davidson 1992; Parkinson and Manstead 1992). Considering that, most of the time, emotional responding is an automatic process, self-reported data is considered to be an unreliable source (Moors 2009). Second, emotion literature in IS research has generally adhered to the classic view of emotion, assuming a limited set of discrete emotions that follow a purely valence-based approach (Storey and Hee Park 2022). Previous work has mostly focused on discrete emotions to investigate: their effect on IT use (Beaudry and Pinsonneault 2010), coping strategies when facing IT failures (Liang et al. 2019), and emotion suppression in online reviews (Hong et al. 2016). However, psychology literature suggests that emotions of the same valence (e.g., gratitude and pride) differ greatly concerning how they affect individuals - from physiological reactions, to judgment, decision-making, and coping strategies (Scherer and Moors 2019; Moors 2016, 2009). Accordingly, there is a need to adopt a more fine-grained view of emotions that recognizes emotions as continuous variables through the dimensions of valence and arousal rather than just discrete emotions (Storey and Hee Park 2022).

The purpose of this study is to address the three mentioned shortcomings in research, these are: (i) the limited theoretical knowledge linking the nature of ITSM with users' emotional responses, (ii) the call for more reliable measures to assess emotions as they occur, and (iii) the current limited view of emotions as discrete entities. Accordingly, this study aims to characterize the nature of ITSM based on the users' emotional response. We adopt a dynamic view of the theory of emotion process focusing on the factors (i.e., IT stimulus) that trigger an emotional episode and drive response differentiation in a continuous fashion. The following question guides our study: *How can we extend the theory of emotion process to characterize the nature of ITSM and emotions?*

As part of the methodological underpinnings of this work, we followed computational grounded theory to propose hypotheses for theoretical inquiry about the link between ITSM and emotions. Our dataset is composed of online reviews about hypertension apps published in an application distribution platform. In this paper, we address how using machine learning methods for health analytics, and online reviews as a source of knowledge can play a defining role in theorizing the nature of ITSM in connection with emotions. This paper proceeds as follows. Section 2 reviews related research concerning ITSM and healthcare analytics using online reviews. Section 3 explains the theoretical underpinning of this work and proposes a hypothesis. Section 4 presents the methodological choices of this study. Section 5 shows the results and section 6 discusses the implications of our findings. Section 7 concludes the paper.

2. **Background**

Hypertension – a highly prevalent disease affecting one in three adults – and the number one risk factor for heart disease and stroke, both of which are leading causes of death (Muntner et al. 2020). Hypertension management requires lifelong self-care of patients by health care professionals for medication dosing (Márquez Contreras et al. 2019), and in many cases, daily support for behavioral changes are needed – however; unlikely to come from health care professionals (Alessa et al. 2018). Accordingly, the responsibility for day-to-day disease management gradually shifts from health care professionals to the individual (Alessa et al. 2018). As the healthcare system is transitioning from reactive care to a more proactive care (Ghose et al. 2022), the use of self-monitoring solutions is recognized.

Mobile health (mHealth) apps, wearable sensors, and devices for home blood pressure monitoring offer great potential in the self-monitoring of hypertension through the involvement of individuals in the self-tracking of their data. mHealth is gaining popularity due to its personalized goal-setting features which have been proven effective in reducing blood pressure, its ability to enhance health literacy through information exchange, and the presence of peer support that transcends time and geographical boundaries (K. Liu, Xie, and Or 2020). Although the topic of self-monitoring of hypertension is not novel (Alessa et al. 2018; Hui et al. 2019; Kazuomi, Harada, and Okura 2022), scholars are concerned that the majority of the studies report that users do not adhere to the apps as intended (Choi et al. 2020; Jakob et al. 2022). Adherence is defined as "the degree to which the user followed the program as it was designed" (Donkin et al. 2011, 2). The nonadherence relative to intended use jeopardizes treatment success which might explain why the World Health Organization recently reported that hypertension is poorly controlled worldwide (World Health Organization (WHO) 2021).

Previous work has reported that user acceptance of hypertension apps decreases over time as a reflection of the users' app-engagement experiences, app technical functionality, and/or design features leading to significant drop outs (Choi et al. 2020; Jakob et al. 2022). Considering that behavioral change for hypertension control requires a long-term commitment, the nonadherence relative to intended use is critical and highlights the necessity to understand the factors that act as barriers to or facilitators of positive user experience (Jakob et al. 2022). To achieve a better understanding of user experience with mHealth, we draw on two streams of research - IT-Enabled Self-Monitoring and health analytics while having the Theory of Emotion Process as our theoretical underpinning. As emotions play a crucial role in shaping experiences, it might be insightful to investigate emotional feedback of patients after using mHealth as a self-monitoring solution.

### 2.1. IT-Enabled Self-Monitoring for Managing Chronic Disease

ITSM is defined as "the use of digital technologies to enable patient SM (i.e., the use of digital technologies to support self-recording of symptoms and behaviors, interpreting the self-recorded data, and adjusting behaviors accordingly)" (Jiang and Cameron 2020, 453). Examples of ITSM include mobile apps, web-based tracking programs, sensing devices, wearable technologies, and insideable devices (e.g., in-body implants, under-skin sensors, or ingestible smart pills) (Jiang and Cameron 2020). For patients with chronic diseases, these strategies help them to manage their health on a day-to-day basis and to increase education and self-awareness. Medical research shows that health apps are effective in the self-monitoring of physiological factors such as decreased blood pressure (Or and Tao 2016) and Body Monitoring Index (BMI) (Blasco et al. 2012; Widmer et al. 2017), cognitive and behavioral factors (e.g., medication adherence) (Anglada-Martínez et al. 2016), and psychological factors (e.g., quality of life) (Ammenwerth et al. 2015). Therefore, ITSM is a promising approach that supports individuals in monitoring factors that are relevant to their health.

To support in the assessment of the experience provided by the ITSM, Hensher et al. (2021) identified ten domains that are important to promote sustained usage of mHealth, these are: (1) Clarity of purpose of the app, (2) Developer credibility, (3) Content/information validity, (4) User experience, (5) User-engagement/adherence and social support, (6) Interoperability, (7) Value, (8) Technical features and support, (9) Privacy/ethics/legal, and (10) Accessibility. The domains were identified and defined based on evaluation of several established mHealth evaluation frameworks such as MARS and uMARS (Stoyanov et al. 2016) and provide unique assessment criteria. In this study we draw on these ten domains as our lexicon, as we argue that the framework provides a blueprint to identify IT stimuli that are emotionally relevant for the user.

### 2.2. Healthcare Analytics & Online Reviews

Healthcare analytics refers to the systematic analysis of health data and patient-generated content to drive decision making for improved healthcare practices. Due to recent technological developments, health analytics methodologies have expanded from econometrics-based models to advanced computational models such as deep learning (Bardhan, Chen, and Karahanna 2020). The possibility to investigate larger datasets using more advanced methods has encouraged scholars to approach traditional research questions with a more granular view.

More recently, researchers have begun to explore deep learning techniques for detecting insights in online reviews as it reflects user feedback which can then be used to better understand users' experience. For instance, previous work has shown topic modeling to be a powerful unsupervised machine learning approach that enables inductive and automated discovery of latent topics in large amounts of texts (Uncovska et al. 2023; Ojo and Rizun 2021;

Lyu, Han, and Luli 2021). This method is widely adopted in user experience studies as a way to investigate what matters to users. A second approach that has recently gained attention from researchers is sentiment analysis, which enables them to investigate users' emotional experiences with products (Lyu, Han, and Luli 2021; Storey and Hee Park 2022). Sentiment analysis includes the labeling of words according to different sentiments, and the training of the model to be able to detect these sentiments in the text. For many years, most studies have used a limited view of emotions as discrete entities, and consequently provided a coarse classification of sentiments (B. Liu 2022; Hong et al. 2016; Liang et al. 2019; Beaudry and Pinsonneault 2010). Recently, scholars have proposed a finer level of sentiments that includes 36 sentiments that were classified according to their valence (negative/positive), arousal (active/passive) and given coordinates on a circumplex model (Storey and Hee Park 2022). Such information helps us to have a more granular understanding of users' emotional responses and provides the necessary framework to analyze the relationship between emotions and an IT artifact.

Online reviews often lack information on basic socio-demographic characteristics of users and are typically large, complex, and unstructured. However, recent studies have demonstrated that these reviews hold promise for knowledge creation when effectively harnessed using advanced health analytics methods. In this paper, we propose that using computational methods of health analytics to investigate online reviews can help us better understand synergies between systems (technology), data (i.e., information), and people (i.e, patients). Considering that both topic modeling and sentiment analysis require an underlying lexicon, we draw on the mHealth evaluation framework (Hensher et al. 2021) as our lexicon for mapping the topics extracted from the topic modeling. Additionally, we use the GoEmotions dataset for fine-grained emotions (Demszky et al. 2020) for sentiment extraction, and the ontology of emotion process (Storey and Hee Park 2022) as our lexicon for sentiment mapping.

## 3. Theoretical Background & Hypothesis Development

By building on previous work (Storey and Hee Park 2022), the theoretical underpinning for the hypotheses include the notion of theory of emotion process (Nico H. Frijda 1986; N. H. Frijda 1988; Scherer and Ellgring 2007) applied to user-generated content for the healthcare industry - i.e., online reviews about ITSM. We developed an overarching framework with which to organize the extant research by adapting the theory of emotion process (Nico H. Frijda 1986). The framework for this study (see Figure 1) aims to expand the theory of emotion process by characterizing the nature of ITSM based on the process of emotion generation and emotion regulation.

The emotion process includes a sequence of components consisting of stimulus, appraisal, valence, arousal, action tendency, and behavior (Nico H. Frijda 1986; Smith and Ellsworth 1985; Scherer 2009). The process begins when an attended situation (e.g., stimulus due to ITSM failure) is interpreted as being central to one's goal including personal (e.g., avoid loss) and social (e.g., help another individual) (Scherer 2001). After the stimulus is appraised to be emotionally significant and relevant, emotions are generated by giving it a valenced meaning, and activating a certain arousal level.

Emotion regulation refers to the effort made by individuals when modulating their emotions in pursuit of their goals (Augustine and Hemenover 2009). The aim of this regulatory goal is either to decrease (downregulate) or increase (upregulate) facets of emotional responding (e.g., intensity).

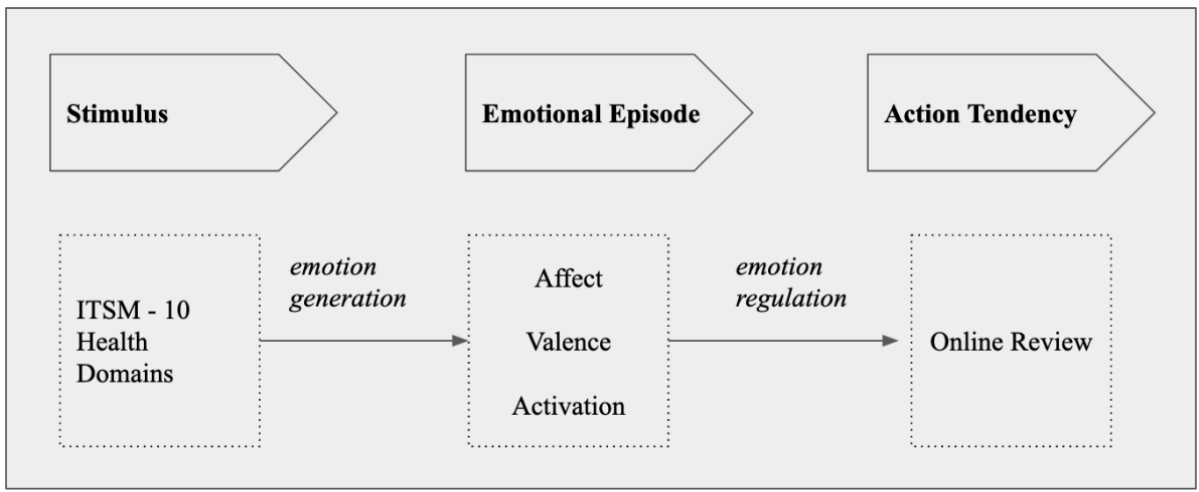

Figure 1 - Theory of emotion process in the context of emotional cues in online reviews about ITSM (adapted from Nico H. Frijda 1986; Storey and Hee Park 2022).

In other words, it happens when individuals try to feel less negative or more positive in the short term. The regulatory target can be to prompt a change within the individual who experiences the emotions (intrinsic) or elicit a change in another person (extrinsic) (Sheppes, Suri, and Gross 2015). Then, an action tendency is produced. In this theory, the emotional components are dominant and act as the primary motivators of actions within the entire emotional sequence.

Although emotion generation and emotion regulation operate within the emotion process and share the same feature of activation of goal, the target of these goals differ. The target of emotion-generation is to emotionally respond to a stimulus, whereas emotion regulation deals with how we manage or modify that emotion once it's present (Gross and Barrett 2011). In this study, we argue that online reviews provide a platform where individuals can fully reveal their emotions and the act of writing a review can also serve as a form of emotion regulation (Xu 2020). In an online context, arousal is unobservable, but can be inferred from the intensity of language (e.g., adjectives, vulgar words) (Park et al. 2019) and the different types of affect (e.g., angry, frustrated) (Russell 1980).

## 3.1 ITSM and emotion generation

A large body of evidence supports the view that a negative IT stimulus leads to the tendency of expressing emotions through online venting or blaming others for the unexpected outcome (Beaudry and Pinsonneault 2010). In particular, IT defects are often described as negative experiences that tend to occur due to the perception of a lack of control over its consequences (Yin, Bond, and Zhang 2014) leading the users to attribute a goal-incongruent event to external sources. Consequently, negative emotions trigger a tendency to acquire temporary situational control through vindictive negative word-of-mouth expressed in online reviews (Gelbrich 2010). For instance, when users experience intense negative emotions because IT is slow or an update compromises user experience (Salo 2020), review writers tend to express those failures to others (e.g., company or other users). The resulting online review expresses the personal relevance that the topic has to the user. In other words, literature has shown that in the case of IT failure, negative emotional discharge is frequently expressed in online reviews after users appraised the incident's relevance to them.

However, in the particular context of ITSM, a great deal of importance is given to the health data that is measured and displayed by the solution as it can mean a difference between life and death (Jones 2014; Chatterjee et al. 2021). Contrary to the use of non-health IT, where information primarily serves auxiliary or supportive functions, the information delivered through ITSM platforms directly influences a person's health and well-being. In the healthcare context, users rely on the information not just for convenience or entertainment, but for critical decisions concerning their health. If that trust is compromised, emotions such as anger, dissatisfaction, and disgust are likely to be experienced by the users. The compromise in utility and the direct risk posed to the user's health can elicit strong negative emotions like fear, anxiety, or anger, reflecting the critical nature of the situation compared to frustrations experienced in regular IT usage. Consistent with the discussion, we argue that the negative stimulus will appraise more negative emotions in the case of unreliable information than system defects. Thus, we hypothesize:

**Hypothesis 1a:** ITSM stimulus caused by the precision of health data influences the number of online reviews more than stimulus caused by technical features.

**Hypothesis 1b:** ITSM stimulus, caused by precision of health data, that are goal-incongruent lead to more negative mean valence than goal-incongruent stimulus concerning technical features.

## 3.2 ITSM and emotion regulation

Emotion regulation consists of attempts to *manage one's own emotions* or *inform decision-makers*. When individuals encounter intense emotions, they often turn to others to help regulate these emotions or the emotions act as motivation for social sharing (Nozaki and Mikolajczak 2022).

Writing reviews offers individuals a platform to share their experiences, thereby providing an avenue to process and manage their emotions. Expressing oneself can act as a cathartic experience, allowing an individual to gain clarity on their feelings (Ullah et al. 2016). When users feel high dissatisfaction, they tend to disclose their emotions to alleviate emotional tension. On the other hand, when users feel high satisfaction, they tend to write about their emotions to celebrate their positive experiences. In addition to acting as a form of catharsis, online reviews are also used to incite change or prevent others from facing similar situations. By writing emotionally-charged reviews, individuals can alert manufacturers about particular issues and potentially instigate improvements (Yin, Bond, and Zhang 2014). Furthermore, such reviews may serve as endorsements for manufacturers to continue to perform well or as encouragement for other users to explore why the review writers are satisfied. Emotion regulation, in this context, is not just about managing one's emotions but also aims to influence decision-makers who are seeking information about the product.

Either as a form of catharsis (emotional release) or product evaluation, online reviews are likely to come from high-intensity emotions. Accordingly, we argue that users who are extremely satisfied or extremely dissatisfied write reviews, while those who felt less extreme emotions may not be motivated to post a review (Ullah et al. 2016). In general, activated emotions possess significant power as motivators for pursuing action. Thus, we hypothesize

**Hypothesis 2:** High activation emotions caused by ITSM stimulus influence the number of reviews more than low activation emotions.

4. **Methodology**

For this study, we followed the recommendations from IS scholars on how to operationalize CGT (Berente and Seidel 2014; Berente, Seidel, and Safadi 2019). Although the main framework of how to operationalize CGT was followed (Berente, Seidel, and Safadi 2019; Berente and Seidel 2014; Nelson 2020), we proposed new techniques for data collection and analysis to be used in the context of analyzing unstructured user reviews about medical apps from the Apple App Store. For instance, while previous studies (Nelson 2020; Ojo and Rizun 2021) applied a three-step approach, our research design comprises four steps, which are explained as follows. Figure 2 illustrates the four-step approach as part of our research design.

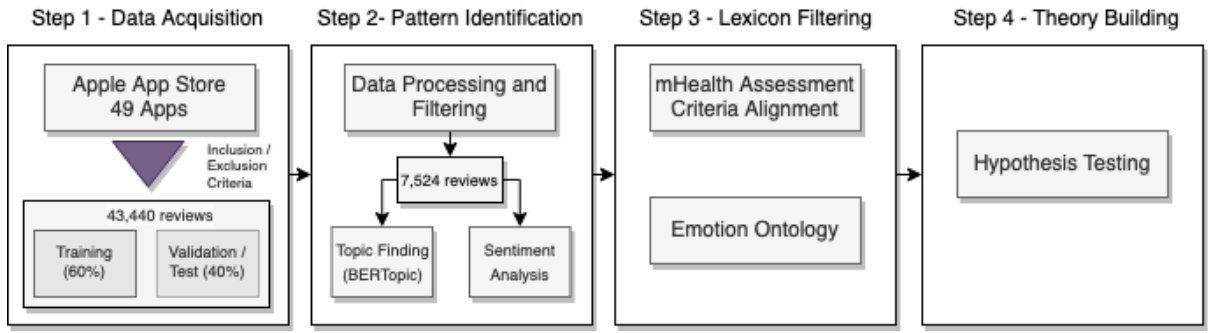

Figure 2 - *We obtain reviews for 49 apps, from which we use 60% as a training set. After preprocessing and filtering, we retain 7,524 reviews for topic modeling and sentiment analysis. We align the identified topics with the mHealth assessment criteria, as well as the sentiment with the Emotion Ontology, which enables us to test our hypotheses.*

### 4.1 Data Acquisition

To enable us to collect the reviews from the Apple App Store, we used the tool Apple Store Analytics which is an open source tool that is available for download. Before using the tool to identify apps that are relevant to our study, we defined a list of inclusion and exclusion criteria as shown in Table 1. The search terms used were "blood pressure" OR "hypertension" (#1). The search looks for the keywords in the description, name, and release notes of the apps on a global level. Moreover, all 175 iTunes countries (#2) and all languages were included (#3). As exclusion criteria, we excluded apps that have less than 100 ratings (#5) due to the very few reviews available. Additionally, we excluded apps with GenreID: 6014 (Games) (#6). As a result, we obtained a list of 196 unique apps. Furthermore, we decided to only include apps that offer connection with wearables (#4) (e.g., Apple Watch, blood pressure monitor). To address that, an additional set of keywords were defined which includes `Apple Health', `Bluetooth', `Connect', `Device', `Record', `Sync', `Watch', `WiFi', `iHealth'. Additionally, we included the names of the most popular blood pressure monitoring devices in the market (e.g.,

Withings and Omron). Considering that having additional keywords is not sufficient to achieve our criteria #4, we decided to manually analyze all pre-selected apps through their App page or Company website which was already listed in the search result.

Our final dataset contains 43,440 processed reviews from 49 applications that either have blood pressure monitoring as their primary function or act as a secondary feature in apps that monitor other conditions, such as diabetes. The average review length (summing the body and the title length) is 36.49 words. In total, 50 languages were detected. 78% of the reviews were written in five languages: English (21630), German (4486), French (2613), Spanish (2563) and Chinese (2056). 75% of the comments are from two continents: Europe and North America.

*Table 1. Inclusion and Exclusion Criteria for App Search.*

| # | **Inclusion Criteria** |
|---|---|
| 1 | Search terms were "blood pressure" OR "hypertension" |
| 2 | All iTunes country stores |
| 3 | All languages |
| 4 | Connection with wearables (e.g., Apple Watch, blood pressure monitor) |
| # | **Exclusion criteria** |
| 5 | Less than 100 global ratings |
| 6 | GenreID: 6014 (Games) |

We used the scikit-learn library (Pedregosa, Varoquaux, and Gramfort 2011) to split the original dataset (43,440 reviews) into training and validation/test sets (60%, 20%/20%). The training set is further pre-processed and used as input to the topic model, as described in the following. Finally, one empty review was excluded from the test set after the preprocessing step. For this work, we use only the remaining instances in the training set and keep the remaining reviews as hold out validation and test data to enable research on supervised machine learning methods in future work.

## 4.2 Pattern Identification

### 4.2.1 Extracting Latent Topics (Topic modeling)

Traditionally, Latent Dirichlet Allocation (LDA) has been used as a topic modeling approach to derive topics and insights from large unstructured text. In this work, however, we used a technique based on BERT (Bidirectional Encoder Representations from Transformers) to leverage the capabilities of pre-trained models and better represent contextual information (Devlin et al. 2018). Following the success of (Alhaj et al. 2022) in analyzing short text and the robust baseline set by (Grootendorst 2022), we selected BERTopic as the technique to analyze the reviews recovered from our Apple Store Analytics tool.

BERTopic uses three phases to produce a topic distribution based on a set of documents—user reviews in our case. First, it creates from each user review a numerical vector representation of the text (i.e. embeddings). Next, it reduces the dimensionality of the embedding vectors using the UMAP algorithm (McInnes, Healy & Melville 2018), and then it clusters the UMAP embeddings with HDBSCAN (Campello, Moulavi, and Sander 2013) to find similar text that composes a topic. The ideal number of topics is initially inferred by BERTopic. In the last step, the importance of words within each topic is defined according to a variation of the Term Frequency-Inverse Document Frequency (TD-IDF) measure named c-TF-IDF (Grootendorst 2022). As seen in Equation 1, this measure treats the documents from each cluster as a single document class $c$ to show the importance (or weight $W$) of term $t$ per class instead of per document.

$$W_{t,c} = tf_{t,c} * \log(1 + A/tf_t) \qquad (1)$$

Before feeding the dataset into BERTopic, we filtered all reviews that had less than 25 words. We used the raw English text to train the BERTopic model and a pre-processed version of the text was applied to improve the results from the c-TF-IDF measure. We removed stop words using the Natural Language Toolkit (NLTK) library (e.g., "the", "and") (Bird, Klein, and Loper 2009) together with a list of common words in the English language and terms such as "app" and "html". Numbers were also removed from the pre-processed texts. Finally, we used lemmatization to reduce word variants to a common base form (e.g., "accessibility" and "accessible" to "access") and selected only the resulting nouns.

After the training and reduction process, 30 topics were identified and were used in the following steps of our research design. To arrive at this number of topics, we tried to maximize the topic coherence using the Normalized Pointwise Mutual Information (NPMI) metric (Bouma 2009) and topic diversity using the percentage of unique words (Dieng, Ruiz, and Blei 2020) in a topic as seen in Figure 3. We chose the number of topics with the highest coherence score and still acceptable diversity. This number of topics lies in the range of 10 to 50 topics, proposed as suitable for human interpretation (Debortoli et al. 2016; Schmitt et al. 2020).

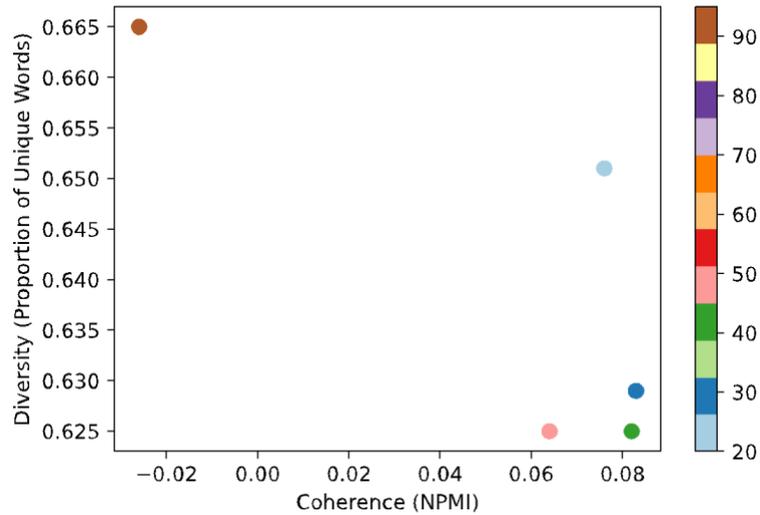

*Figure 3. This figure depicts the NPMI coherence score in the x axis and the topic diversity in the y axis. The original output of BERTopic was 95 topics which were later reduced to 30.*

### 4.2.2 Extracting Sentiments (Sentiment Analysis)

After all pre-processing and filtering steps, the final dataset contained 7,524 reviews that were subject to sentiment analysis. Our method is based on the GoEmotions dataset and emotion taxonomy (Demszky et al. 2020). The taxonomy is based on psychological research (Cowen and Keltner 2017) and includes 28 fine-grained emotion categories (including a neutral category), which cover positive, negative, and ambiguous emotions. A dataset of 58k diverse social media texts was manually annotated according to these categories and made publicly available.

To predict emotions in our dataset, we use a model following the standard BERT-based text classification architecture, which was trained on the GoEmotions dataset and published on the Hugging Face Hub[1]. The input of the model is an arbitrary text snippet, and the output a probability distribution over 28 categories. We apply the model to all review texts in the training dataset and normalize the resulting probabilities to have zero mean and unit variance. We refer to the resulting normalized values as *emotion scores*.

---

[1] https://huggingface.co/bhadresh-savani/bert-base-go-emotion

### 4.3. Lexicon Filtering

#### 4.3.1. Mapping Latent Topics to the mHealth Evaluation Framework

The third step of our research design, lexicon filtering, comprises the interpretation and sense-making of the valid topics that the BERTopic revealed upon the training set. To perform this step, the following analysis was carried out: (1) the definition associated with each of the Mobile App Evaluation Framework (Hensher et al. 2021) were reviewed; (2) two scientists independently mapped each of the 30 identified topics to exactly one dimension of the framework and discussed until common agreement was achieved.

#### 4.3.2. Mapping Sentiments to the Circumplex Model of the Ontology of Emotion Process

To obtain valence and activation values for each emotion in GoEmotions, we manually mapped the categories to the affected categories of the emotion ontology (Storey and Hee Park 2022). The results are shown in Figure 4 and Appendix. A.

A direct mapping is possible for most categories. The category "Fear" is absent from Storey and Park's model, but can be obtained from the prior circumplex model (Scherer 2005). The "Approval" category is absent from both models. However, from the analysis performed by (Demszky et al. 2020), we can infer that approval most likely corresponds to "Contentment" in (Storey and Hee Park 2022) model.

.

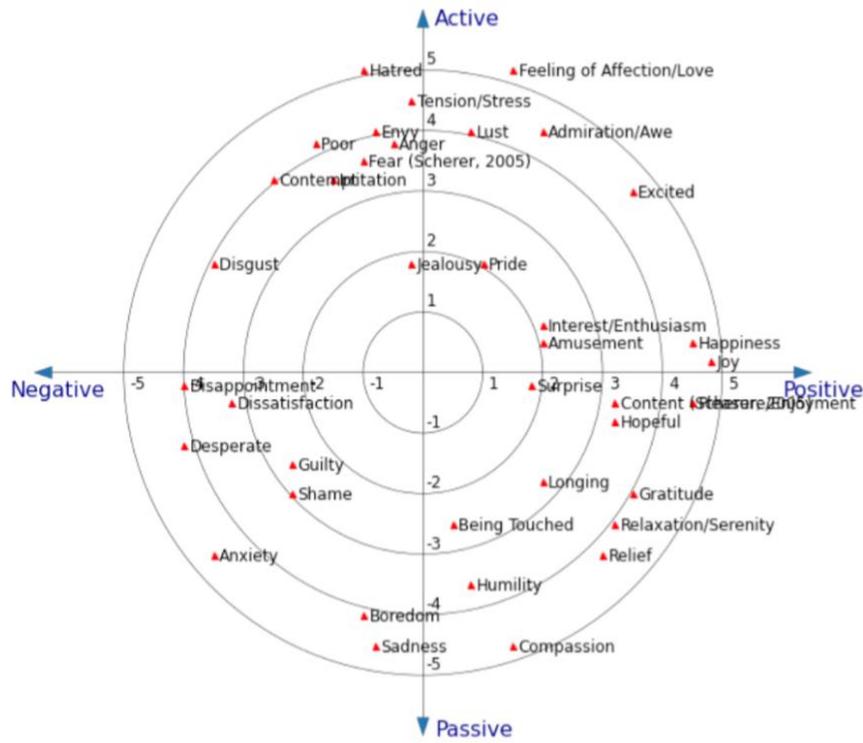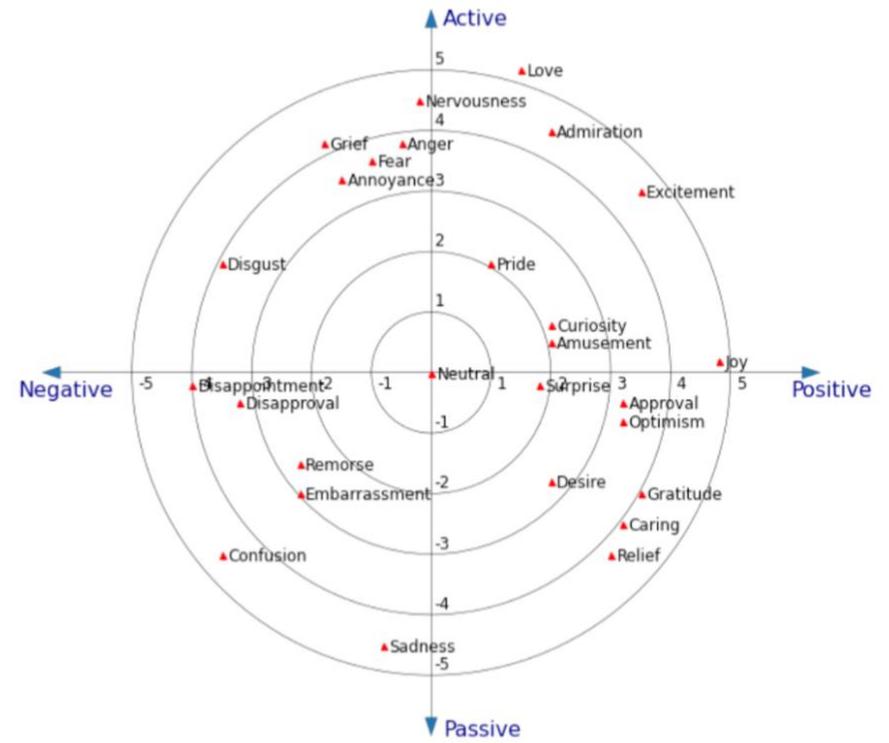

*Figure 4. Reconstruction of the Emotion Ontology introduced by Storey and Park and Mapping of GoEmotions categories to the same coordinate system of*

Further, we assume that "Neutral" corresponds to a value of 0 in both the valence and activation dimension. "Realization" is the only category from the *valence and activation.*

Further, we assume that "Neutral" corresponds to a value of 0 in both the valence and activation dimension. "Realization" is the only category from the GoEmotions taxonomy that we cannot reliably map to any coordinate in the circumplex model, and is therefore not included in further analysis[2].

Using the obtained mapping, each text document (i.e., review) can be assigned a coordinate in the circumplex model. We do so by calculating an average of activation and valence values for all identified sentiments in these emotions, weighted by their normalized emotion score (see Sect. 3.2.2). Due to the normalization, the expected value for both valence and activation across all documents is 0, i.e., neutral.

To identify significant deviations from the neutral mean (0,0) for each topic, we consider the subset of reviews where a topic is most dominant and thus calculate the mean valence and activation for each topic. We perform a statistical significance test ($t$-test) on the difference between the mean valence and activation from the expected mean 0 and apply a Bonferroni correction to account for multiple testing across all topics. We consider corrected $p$-values > 0.05 as significant

### 4.4 Theory Building

The overarching framework for this study was theoretically built, but the operationalization from its variables was derived from the data. For instance, the list of the ITSM stimulus was theoretically derived from the mHealth evaluation framework (Hensher et al. 2021), but its operationalization (i.e., the decision of which domain would be tested) emerged after an initial manual analysis of the data. The variables were then used for hypothesis testing. These are:

---

[2]The complete lexicons from Storey and Park could potentially help with this mapping, but after request, they were not available.

**ITSM stimulus:** Based on the mHealth evaluation framework, the variable ITSM stimulus was operationalized using the domains of "Content/Information Validity" and "Technical Features and Support". "Content/Information Validity" refers to the accuracy of the information in the health app, whereas the latter refers to defects, errors, bugs and quality after updates. It is important to note that although "Interoperability" is also a technical topic, it was not included in our operationalization as our intent was to investigate stimulus referred to specific features of the app and not how the app and wearable device work together seamlessly.

**Online Reviews:** Reviews were downloaded using the Apple Store Analytics based on inclusion and exclusion criteria. The reviews were categorized using topic modeling and subjected to sentiment analysis for the detection of affect. Subsequently, pattern matching was employed to map the detected affect to its corresponding valence and activation.

**Affect, Valence and Activation:** Sentiment (i.e., affect) was measured using the GoEmotions dataset (Demszky et al. 2020) and then mapped with the Ontology of Emotions (Storey and Hee Park 2022). The mapping enabled us to allocate different sentiments along the dimensionality of both valence and activation. The three components were treated as individual variables.

## 5. Results

### 5.1    Mapping Topics to the Health App Evaluation Framework

Table 2 presents a detailed overview of the 30 topics that we identified mapped with the dimensions of the app evaluation framework (Hensher et al. 2021), and Table 5 lists the hypotheses tested. Below we present the most interesting findings from each dimension following a ranking of the number of reviews.

Although the dimension "Content/Information Validity" (1498 reviews) only has two topics, it is the most discussed dimension from our dataset. In particular, the topic "Body Mass Index Monitoring (BMI)" (T0) has 1078 reviews, being the topic with the most reviews in our dataset. Reviews within this dimension are concerned with accuracy in BMI and heart rate measures (T5). "Content/Information Validity" refers to readability, credibility, characteristics, quality, and accuracy of the information in the app. The second dimension more discussed was "Interoperability" (1433 reviews). Interoperability refers to the data sharing and data transfer capabilities of the apps. Our dataset indicates that relevant topics for the users include issues related to synchronization when transferring data from blood pressure monitoring devices to the app (T7,T26). Moreover, many reviews seek for a functional integration between the app and smartwatches (T7, T26,T14,T15), and between the app and the Apple Health App (T9). While most topics indicate more general concerns about connectivity issues, one topic refers specifically to issues in tracking when syncing pedometers and the app (T8). The third most frequently discussed dimension was Value (N=1391) which refers to perceived benefits and advantages associated with the use of the app. Topics that are discussed in this dimension indicate that users value having an option to export the measurement of health data in PDF (T23) so that it can be shared with health care professionals (T6). Moreover, the export function enables users to have a health diary (T1).

"Technical Features and Support" (1058 reviews) refers to defects, errors, quantity and timely updates, and technical support and service quality provided within the app. The reviews within this dimension indicate user concerns about defects in the system, such as the app crashing very often (T4), several bugs in the app and smartwatch (T29), and problems with registration and login to the app (T19). Additionally, users voiced concerns about the impact of the system update on the app in terms of changes in the app interface (T21, T25), and new synchronization issues due to updates (T20).

**Table 2.** Topic modeling Mapped to the Evaluation App Framework

| Topic | Theme | Domain |
|---|---|---|
| Clarity when canceling subscription (T24) | Subscription Plan | Clarity of the purpose of the app |
| Transfer of a product ownership to another company (T3) | Brand Image | Developer credibility |
| Body Mass Index monitoring (T0) | Presentation of content & Accuracy | Content/information validity |
| Heart rate notification/alert (T5) | Accuracy | Content/information validity |
| Glucose monitoring (T2) | Functionality | User experience |
| Sleep time recognition (T11) | Functionality | User experience |
| Nutrition planning & mentoring (T22) | Useful feedback & Personalized Experience | User experience |
| Simple interface and navigation (T12) | Ease of use, Design and Functionality | User experience |
| Blood pressure monitoring devices (T7) | Data Transfer (Synchronization) | Interoperability |
| Blood pressure monitoring devices (Vendor Specific) (T26) | Data Transfer (Synchronization) | Interoperability |
| Step tracking (T18) | | |
| Apple health integration (T9) | | |
| Smart watch synchronization (T14) | | |
| Apple watch compatibility (T15) | | |
| Bluetooth pairing (T8) | | |
| Reliable data for health diary (T1) | Perceived usefulness | |
| Data report export for sharing (e.g., with the doctor) (T6) | | |
| Report overview in PDF (T23) | | Value |
| Vendor-specific device (T16) | | |
| Graphical representation (T28) | | |
| Stress and heart rate variability (T10) | Increase knowledge/Awareness | |
| System crash problems (T4) | System defect | |
| Bugs in the app and watch (T29) | | Technical features and support |
| Registration and login (T19) | | |
| Access to readings (T27) | System defect & Customer Support | |
| Synchronization issues (T20) | System Update | |
| Impact of new app interface on user satisfaction (Vendor Specific) (T21) | | |
| Impact of new app interface on user satisfaction (T25) | | |
| Personal data collection and sharing with third parties (T17) | Privacy | Privacy/security/ethical/legal |
| Language support (T13) | Availability | Accessibility |

"User Experience" (1057 reviews) refers to the experience of using an app in terms of user-friendliness, design features, functionalities, and ability to provide a personalized experience. Within this dimension, the most mentioned functionalities were glucose monitoring (T2) and sleep time recognition (T22). Additionally, the reviews indicate a need for a personalized experience through mentoring as part of a nutrition program (T22). Preferences for a simple interface and navigation (T12) were also found in the data. "Developer Credibility" (444) refers to the transparency of the app development and testing processes, and accountability and credibility of the app developer, affiliations, and sponsors. Reviews categorized within this dimension indicate concerns over when a company that has a damaged brand image takes over the product of a company with good branding (T3).

Less frequently discussed topics are related to "Accessibility" (183 reviews), "Privacy/Security/Ethical/Legal" (141 reviews), and "Clarity of the Purpose of the App" (104 reviews). Accessibility refers to the ability of the app to capture a wider audience, which is represented in our dataset by availability of language support (T13). Privacy/Security/Ethical/Legal pertains to data protection and legalities of the health app concerning whether the health app adhere to guidelines. Within this dimension, reviews indicate concerns over what type of data is collected from the app and whether the data is shared with third parties (T17). Finally, "Clarity of the Purpose"of the App is concerned with a clear statement of the intended purpose of the app being provided. Our findings indicate that users seek more clarity concerning subscription plans, in particular to know whether they will be charged after canceling a subscription (T24).

Interestingly, we did not find indications for topics related to the dimension of "User Engagement/Adherence" and "Social Support" which refers to the extent of how apps maintain user retention using functionalities such as gamification, forum, behavior technique and social support. It is surprising because using intervention techniques for behavioral change and peer support are fundamental for self-management (Hui et al. 2019; Fallon et al. 2021).

## 5.2    Emotion generation as a detector of ITSM stimulus relevance

Hypothesis 1A proposed that ITSM stimulus caused by the precision of health data influences the number of online reviews more than stimulus caused by technical features. Consistent with the hypothesis, "Content/Information Validity" (1498 reviews) was the domain that influenced the number of reviews the most, followed by "Interoperability" (1433 reviews), "Value" (N=1391), and finally "Technical Features and Support" (1058 reviews). *Hence, Hypothesis 1A was supported.*

Due to its pervasive nature, many individuals rely on the information provided by the app to understand and make decisions about their condition. Any deviation from factual accuracy can lead to the mismanagement of the condition, which can be life-threatening. Thus, the relevance of having access to precise content is deemed to be the most relevant topic in the context of ITSM. By using machine learning techniques, we can identify cues in online reviews concerning the users' emotion generation and use the insights as a detector of relevance concerning ITSM stimulus.

Hypothesis 1B proposed that ITSM stimulus, caused by precision of health data, that are goal-incongruent lead to more negative mean valence than goal-incongruent stimulus concerning technical features. Inconsistent with the hypothesis, the mean valence of "Technical features and support" is more negative than "Content/Information validity". *Hence, Hypothesis 1B was rejected.* Table 3 provides data on the valence for both mHealth domains. See Appendix B for the valence of all domains.

Table 3 - Mean valence of reviews with topics related to the mHealth domains "Technical features and support" and "Content/Information validity". Both values are significantly different from the expected mean 0 (p < 0.05)

| mHealth Domain | Valence (mean) | p |
|---|---:|---:|
| Technical features and support | - 0.263 | <u><0.001</u> |
| Content/Information validity | - 0.071 | <u>0.009</u> |

Interestingly, our results suggest that while precision of content is identified as the most relevant topic, negative stimuli related to it do not trigger strong emotional responses. Although further research is required, the finding suggests that the ITSM stimulus related to precision of content might be a more complex, cognitive-driven interaction than a purely emotional one. When inaccuracies are encountered, users' primary response might be cognitive dissonance rather than an emotional outburst. In this way, the nature of the users' engagement with content-related stimulus that is goal incongruent might have a cognitive prioritization over emotion.

### 5.3 Emotion regulation as a cause of action tendency in the context of ITSM

Hypothesis 2 proposed that high activation emotions caused by ITSM stimulus influence the number of reviews more than low activation emotions. While the results are inconclusive for most of the topics, our data indicated two topics that are significantly passive and two topics that are significantly active, which are listed in Table 4. The two topics that elicit passive emotions (low activation) are concerned with "Glucose Monitoring (T2)", and "Smart Watch Synchronization (T14)". The two topics that elicit activated emotions are concerned with "Transfer of a product ownership to another company (T3)", and "Impact of new app interface on user satisfaction (Vendor Specific)" (T21). *Hence, Hypothesis 2 was rejected.*

Table 4 - Topics with mean activation levels significantly different from 0 (p < 0.05)

| mHealth Domain | Topic | Activation | p |
|---|---|---|---|
| User Experience | Glucose Monitoring (T2) | -0.073 | 0.045 |
| Developer Credibility | Transfer of a product ownership to another company (T3) | 0.143 | <0.001 |
| Interoperability | Smart Watch Synchronization (T14) | -0.142 | 0.001 |
| Technical features and support | Impact of new app interface on user satisfaction (Vendor Specific)" (T21) | 0.337 | <0.001 |

It is interesting to note that the only two topics that evoke high activation emotions refer to brands. Topic T3 refers to the users' dissatisfaction when the company of a product that they use is sold to a company that they dislike, whereas Topic T21 refers to users' dissatisfaction after a trusted company releases updates that negatively affect the user experience. Although the literature has highlighted several times (Gelbrich 2010) that dissatisfaction related to technical features tends to evoke high emotions in users, surprisingly, topic T14 which refers to synchronization issues is found to evoke low activation emotions. Even more surprising considering that interoperability is fundamental for displaying the results in the app from the blood pressure monitoring device. Finally, the feature of "Glucose Monitoring" tends to not evoke high arousal levels in the user in the context of hypertension apps.

Since most of the topics did not yield significant activation results, we performed the analysis on the level of reviews. As a result, 4285 have negative activation and 3239 have positive activation. The median is -0.07 (slightly more passive) - most of the low activation reviews are very close to the mean 0. Figure 5 illustrates the distribution of the reviews according to their valence.

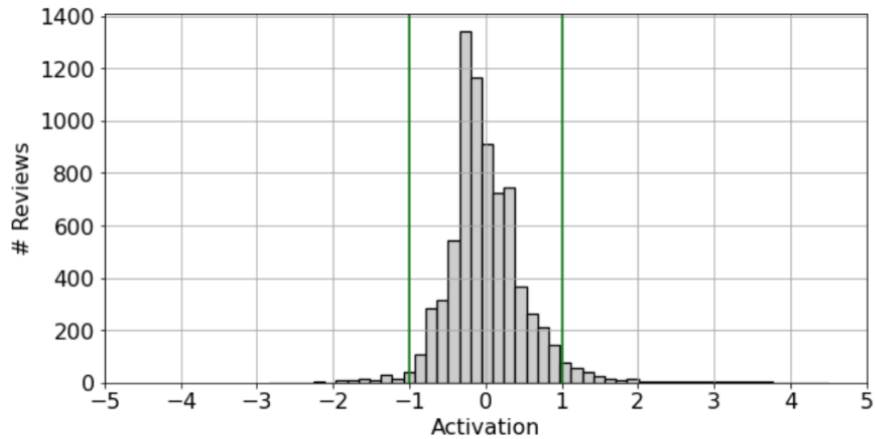

Figure 5 - Histogram of activation levels on the level of individual reviews. Most reviews have activation levels close to 0, although the distribution is slightly skewed, i.e., there is a longer tail of reviews with high positive activation levels.

However, if we follow the recommendations from the Ontology of Emotion Process (Storey and Hee Park 2022), the range [-1,1], which is indicated by the green line, should be considered neutral. By following that definition, there are 106 reviews with an activation < -1 and 247 with > 1. Although the number of reviews was not our level of analysis, but instead the number of topics, it is interesting to see the difference in the results when we consider the range [-1,1] neutral or not.

By following the Ontology of Emotion Process to analyze the data on the level of reviews, we found that most reviews fall into the neutral range. When looking at the reviews that are outside the neutral range, there are more reviews with higher activation than reviews with low activation. At the same time, if we consider that there are no neutral sentiments, there are more low activation reviews than high activation reviews.

Table 5 - Overview of supported and rejected hypotheses.

| Component of the theory of emotion process | Hypothesis | Result |
|---|---|---|
| Emotion Generation | H1a: ITSM stimulus caused by the precision of health data influences the number of online reviews more than stimulus caused by technical features. | Supported |
| | H1b: ITSM stimulus, caused by precision of health data, that are goal-incongruent lead to more negative mean valence than goal-incongruent stimulus concerning technical features. | Rejected |
| Emotion Regulation | H2: High activation emotions caused by ITSM stimulus influence the number of reviews more than low activation emotions. | Rejected |

## 6. Discussion

This study characterizes the nature of ITSM based on the users' emotional response to ITSM stimulus that is regulated through online reviews. We employed machine learning methods to predict the occurrence of emotions in online reviews about ITSM for hypertension. By using a dynamic view of the theory of emotion process, we use the concepts of stimulus, valence, arousal, affect, and action tendency to explain emotion generation and emotion regulation in the context of ITSM. The findings reported in this paper have significant implications for research at the intersection of IS and healthcare.

We expanded previous work (Jiang and Cameron 2020; Chatterjee et al. 2021) in ITSM by investigating the causal mechanisms of ITSM stimulus and users' emotional expressions. Our findings show that existing theories yet do not account for the very specific context that is the use of self-monitoring solutions for chronic care. A clear contradiction found in the literature is that although IT and ITSM have been treated as if they were the same, studies have highlighted clear differences between them in terms of the importance of information. A key finding is the critical nature of health information that is collected and displayed by these solutions, which characterizes the ITSM artifact as a purposive system that seeks multiple goals through multiple paths (multifinality and equifinality). Additionally, we provided a novel perspective on emotion generation and emotion regulation by taking a dynamic view of emotion (Nico H. Frijda 1986; Storey and Hee Park 2022) and investigating those concepts in the context of ITSM. We propose to see emotion generation as a detector of relevance of ITSM stimulus, and online reviews as mechanisms for regulating those emotions. By doing that, we discuss how recent technological advances are offering new venues for users to express their emotions. Finally, while emotion literature in IS research has followed a purely valenced-approach, this study demonstrates the ability to reliably detect fine-grained valence and arousal levels by using machine learning models. By examining both the valence and arousal aspects of emotions, we can achieve a deeper understanding of users' emotional states. This knowledge can then guide approaches to enhance user experience, address problems efficiently, and respond to user feedback aptly. For instance, previous work (Jakob et al. 2022; Donkin et al. 2011)) have highlighted user experience as a reason for ITSM discontinuation, however; little is known about how emotions of the user towards particular company brands can lead to ITSM discontinuation.

Our study contributes to theory by advancing the theory of emotion process to account for the very unique context that is the self-monitoring IT artifact for healthcare. Methodologically, it presents how to operationalize computational grounded theory to predict users' emotions beyond self-reports. For practice, the use of our methods for detecting ITSM stimulus and sentiment analysis can support companies in product development, customer relationship and brand management.

## 7. Conclusion

This research has extended the theory of emotion process to characterize the nature of ITSM based on the users' emotional response. While previous studies have focused on detecting discrete human emotions, we set out to investigate the link between ITSM stimulus and continuous valence and arousal levels. By taking a dimensional view of emotions, our findings present a novel perspective for traditional components of emotion theories - emotion generation and emotion regulation. Our findings indicate emotion generation to be seen as an indicator of relevance of ITSM stimulus. Moreover, writing online reviews can be understood as a psychological way to regulate the emotions triggered by these stimuli. This study offers a comprehensive understanding of emotional responses in the context of IT-enabled self-monitoring for chronic care, both in terms of their generation and their expression.

**Appendix A:** Mapping GoEmotions (Demszky et al. 2020) and Emotion Ontology (Storey and Hee Park 2022) and their coordinates on the circumplex model of Emotion Process

| Emotion Ontology (Storey and Hee Park 2022) | GoEmotions (Demszky et al. 2020) | Valence | Activation |
|---|---|---:|---:|
| Feeling of Affection/Love | Love | 1.5 | 5.0 |
| Admiration/Awe | Admiration | 2.0 | 4.0 |
| Pride | Pride | 1.0 | 1.8 |
| Joy | Joy | 4.8 | 0.2 |
| Interest/Enthusiasm | Curiosity | 2.0 | 0.8 |
| Amusement | Amusement | 2.0 | 0.5 |
| Excited | Excitement | 3.5 | 3.0 |
| Lust | - | 0.8 | 4.0 |
| Happiness | - | 4.5 | 0.5 |
| Tension/Stress | Nervousness | -0.2 | 4.5 |
| Anger | Anger | -0.5 | 3.8 |
| Disgust | Disgust | -3.5 | 1.8 |
| Irritation | Annoyance | -1.5 | 3.2 |
| Poor | Grief | -1.8 | 3.8 |
| Fear (Scherer, 2005) | Fear | -1.0 | 3.5 |
| Contempt | - | -2.5 | 3.2 |
| Envy | - | -0.8 | 4 |
| Jealousy | - | -0.2 | 1.8 |
| Hatred | - | -1.0 | 5 |
| Gratitude | Gratitude | 3.5 | -2 |
| Relaxation/Serenity | Caring | 3.2 | -2.5 |
| Relief | Relief | 3 | -3 |
| Longing | Desire | 2 | -1.8 |
| Surprise | Surprise | 1.8 | -0.2 |
| Hopeful | Optimism | 3.2 | -0.8 |
| Content (Scherer, 2005) | Approval | 3.2 | -0.5 |
| Compassion | - | 1.5 | -4.5 |
| Humility | - | 0.8 | -3.5 |
| Pleasure/Enjoyment | - | 4.5 | -0.5 |
| Being Touched | - | 0.5 | -2.5 |
| - | Realization | - | - |
| Disappointment | Disappointment | -4 | -0.2 |
| Dissatisfaction | Disapproval | -3.2 | -0.5 |
| Shame | Embarrassment | -2.2 | -2 |
| Anxiety | Confusion | -3.5 | -3 |
| Sadness | Sadness | -0.8 | -4.5 |
| Guilty | Remorse | -2.2 | -1.5 |
| Desperate | - | -4 | -1.2 |
| Boredom | - | -1 | -4 |
| - | Neutral | 0 | 0 |

**Appendix B** - List of topics ranked by lowest valence

| Topic | Dimension | Valence | Activation | p (Valence, corrected) |
|---|---|---|---|---|
| Impact of new app interface on user satisfaction | Technical features and support | -0.51 | **0.15** | <0.001 |
| Registration and login | Technical features and support | -0.50 | **0.00** | <0.001 |
| Impact of new app interface on user satisfaction (Vendor Specific) | Technical features and support | -0.42 | **0.34** | <0.001 |
| Synchronization Issues | Technical features and support | -0.39 | **-0.04** | <0.001 |
| Blood Pressure Monitoring Devices | Interoperability | -0.39 | **0.07** | <0.001 |
| Bluetooth Pairing | Interoperability | -0.38 | **-0.02** | <0.001 |
| Smart Watch Synchronization | Interoperability | -0.37 | **-0.14** | <0.001 |
| Clarity when cancelling subscription | Clarity of the purpose of the app | -0.35 | **0.05** | <0.001 |
| Personal data collection and sharing with third parties | Privacy/Security/Ethics/Legal | -0.34 | **0.08** | <0.001 |
| Transfer of a product ownership to another company | Developer Credibility | -0.34 | **0.14** | <0.001 |
| Apple Health Integration | Interoperability | -0.32 | **-0.09** | <0.001 |
| Access to Readings | Technical features and support | -0.27 | **0.05** | 0.063 |
| Body Mass Index Monitoring | Content/information validity | -0.24 | **-0.04** | <0.001 |
| Blood Pressure Monitoring Devices (Vendor specific) | Interoperability | -0.23 | **0.07** | 0.041 |
| Step Tracking | Interoperability | -0.22 | **-0.01** | 0.034 |
| System Crash Problems | Technical features and support | -0.09 | -0.04 | 0.975 |
| Bugs in the App and Watch | Technical features and support | -0.07 | -0.16 | 1.000 |
| Vendor-Specific Device | Value | -0.04 | 0.01 | 1.000 |
| Graphical representation | Value | 0.11 | 0.06 | 1.000 |
| Language support | Accessibility | 0.11 | -0.13 | 1.000 |

| Feature | Category | Value | Δ | p |
|---|---|---|---|---|
| Apple Watch Compatibility | Interoperability | 0.30 | **-0.01** | <0.001 |
| Glucose monitoring | User experience | 0.31 | **-0.07** | <0.001 |
| Sleep Time Recognition | User experience | 0.32 | **-0.08** | <0.001 |
| Heart Rate Notification/Alert | Content/information validity | 0.37 | **0.00** | <0.001 |
| Stress & Heart Rate Variability | Value | 0.38 | **0.14** | <0.001 |
| Data Report Export for Sharing (e.g. with the doctor) | Value | 0.41 | **0.03** | <0.001 |
| Reliable Data for Health Diary | Value | 0.42 | **-0.04** | <0.001 |
| Simple Interface and Navigation | User experience | 0.45 | **0.00** | <0.001 |
| Report Overview in PDF | Value | 0.56 | **0.02** | <0.001 |
| Nutrition Planning & Mentoring | User experience | 0.66 | **0.10** | <0.001 |